\documentclass[dvipsnames,twocolumn]{aastex631}
\usepackage{graphicx}   
\usepackage{lineno}
\usepackage{bm}
\usepackage{paralist}

\shorttitle{Cosmology with few galaxies}
\shortauthors{C. Chawak et al.}

\graphicspath{{./}{figures/}}

\begin{document}

\title{Cosmology with multiple galaxies}

\author{Chaitanya Chawak}
\affiliation{Indian Institute of Science Education and Research (IISER) Tirupati, Tirupati-517507, India}

\author[0000-0002-4816-0455]{Francisco Villaescusa-Navarro}
\affiliation{Center for Computational Astrophysics, Flatiron Institute, 162 5th Avenue, New York, NY, 10010, USA}
\affiliation{Department of Astrophysical Sciences, Princeton University, Peyton Hall, Princeton NJ 08544, USA}

\author{Nicol\'{a}s Echeverri Rojas}
\affiliation{Instituto de Fisica, Universidad de Antioquia, A.A.1226, Medellin, Colombia}

\author[0000-0001-7899-7195]{Yueying Ni}
\affiliation{Harvard-Smithsonian Center for Astrophysics, 60 Garden Street, Cambridge, MA 02138, US}
\affiliation{McWilliams Center for Cosmology, Department of Physics, Carnegie Mellon University, Pittsburgh, PA 15213, US}

\author[0000-0003-1197-0902]{ChangHoon Hahn}
\affiliation{Department of Astrophysical Sciences, Princeton University, Peyton Hall, Princeton NJ 08544, USA}

\author[0000-0001-5769-4945]{Daniel Anglés-Alcázar}
\affiliation{Center for Computational Astrophysics, Flatiron Institute, 162 5th Avenue, 
New York, NY, 10010, USA}
\affiliation{Department of Physics, University of Connecticut, 196 Auditorium Road, U-3046, 
Storrs, CT, 06269, USA}

\correspondingauthor{Chaitanya Chawak}
\email{chaitanyachawak001@gmail.com}

\begin{abstract}
Recent works have discovered a relatively tight correlation between $\Omega_{\rm m}$ and properties of individual simulated galaxies. Because of this, it has been shown that constraints on $\Omega_{\rm m}$ can be placed using the properties of individual galaxies while accounting for uncertainties on astrophysical processes such as feedback from supernova and active galactic nuclei. In this work, we quantify whether using the properties of multiple galaxies simultaneously can tighten those constraints. For this, we train neural networks to perform likelihood-free inference on the value of two cosmological parameters ($\Omega_{\rm m}$ and $\sigma_8$) and four astrophysical parameters using the properties of several galaxies from thousands of hydrodynamic simulations of the CAMELS project. We find that using properties of more than one galaxy increases the precision of the $\Omega_{\rm m}$ inference. Furthermore, using multiple galaxies enables the inference of other parameters that were poorly constrained with one single galaxy. We show that the same subset of galaxy properties are responsible for the constraints on $\Omega_{\rm m}$ from one and multiple galaxies. Finally, we quantify the robustness of the model and find that without identifying the model range of validity, the model does not perform well when tested on galaxies from other galaxy formation models.
\end{abstract}


\section{Introduction} 
\label{sec:intro}

Some of the most fundamental questions we can ask in cosmology are: What are the components that make up the Universe? How much does each component contribute? We now know that the Universe should be made up of at least three main components: 1) baryons, representing all the substances and materials we know, 2) dark matter, some fundamental particle that interacts with baryons mostly (perhaps uniquely) through gravity, and 3) dark energy, a mysterious substance (perhaps a property of the vacuum) responsible of the recent acceleration of the Universe. From cosmological data, we believe these three components represent roughly $5\%$, $25\%$, and $70\%$ of the current energy content of the Universe. 

Parameters such as $\Omega_{\rm b}$ and $\Omega_{\rm m}$ represent the fraction of the Universe's energy content in terms of baryons and baryons plus dark matter, respectively. Determining them is important to learn about the nature and properties of dark matter and also to learn about the growth rate of the Universe \citep{Huterer_2023}. There are many different methods to infer these parameters, from studying the properties of the cosmic microwave background anisotropies to the spatial distribution of galaxies. Recently, \cite{cosmo1gal} claimed that a tight relation between $\Omega_{\rm m}$ and the properties of individual galaxies are present in galaxies from state-of-the-art hydrodynamic simulations. The relationship is present even when varying the value of astrophysical parameters controlling the efficiency of supernova and active galactic nuclei (AGN) feedback. \cite{Echeverri_2023} reached the same conclusion when using galaxies generated with a different hydrodynamic and subgrid physics model.

\cite{cosmo1gal} discussed that such a relation might be due to the existence of a low-dimensional manifold where galaxy properties reside. In this view, changing $\Omega_{\rm m}$ modifies the location of the galaxies in that manifold differently than changing the efficiency of astrophysical processes. For instance, increasing the value of $\Omega_{\rm m}$ while keeping $\Omega_{\rm b}$ fixed will increase the overall dark matter content of the Universe. That excess will enhance the dark matter content of galaxies, affecting their density, star formation rate, metallicity...etc. On the other hand, feedback can also affect some of these properties, but it is unlikely that it will significantly affect the dark matter content of most galaxies.

\cite{cosmo1gal} argued that knowing the location of one point in the manifold is enough to characterize it, and therefore, with one single galaxy, it is possible to infer the value of $\Omega_{\rm m}$. We note that \cite{cosmo1gal} and \cite{Echeverri_2023} showed that $\Omega_{\rm m}$ can be inferred with a $\sim10\%$ precision based on the properties of a single galaxy, perhaps indicating that the manifold should have some intrinsic width associated with it. However, by using multiple galaxies, it should also be possible to infer the value of cosmological and astrophysical parameters by characterizing the impact on galaxy statistics like the stellar mass function. Recently, \cite{Busillo_2023} have shown that galaxy scaling relations are sensitive to both cosmology and astrophysics and derived constraints on those from real data \citep[see also ][for the impact on the star-formation rate history and the stellar mass function]{Jo_2022}. In this work, we thus ask ourselves how well we can infer cosmological parameters if we only have a few galaxies. Due to computational constraints, we limit our analysis to less than ten galaxies. We note that using properties from galaxies directly instead of summary statistics enables our models to search through all potential summary statistics and cross-correlations.

In this paper, we show that using more than one galaxy increases the precision of the models trained to infer $\Omega_{\rm m}$, but at the same time, allows the models to infer other parameters that were unconstrained when using a single galaxy. To carry out our analysis, we made use of thousands of state-of-the-art hydrodynamic simulations from the Cosmology and Astrophysics with MachinE Learning Simulations (CAMELS) project\footnote{\url{https://www.camel-simulations.org/}} \citep{CAMELS, CAMELS_public, CAMELS-Astrid}. To check that our results do not hold just for galaxies generated by a particular code, we perform our analysis using simulations run with three codes that employ different subgrid physics models: 1) AREPO+IllustrisTNG, 2) GIZMO+SIMBA, and 3) MP-Gadget+Astrid. 

This paper is organized as follows. We present the data we use and the machine learning algorithms we employ in Section \ref{sec:methods}. In Section \ref{sec:results}, we present the main results of our analysis. Finally, we summarize the takeaways and conclude in Section \ref{sec:summary}.


\section{Methods} 
\label{sec:methods}

In this section, we first describe the data we use for this work. We then explain the machine learning algorithms we employ to analyze the data and outline the metrics we utilize to quantify the accuracy and precision of our models.

\subsection{Data}
\label{subsec:data}

In this paper, we train neural networks to infer the value of cosmological and astrophysical parameters using the internal properties of simulated galaxies. These galaxies come from state-of-the-art hydrodynamic simulations of the CAMELS project \citep{CAMELS}. 

All simulations follow the non-linear evolution of $256^3$ dark matter plus $256^3$ initial fluid elements from $z=127$ down to $z=0$ in a cubic periodic volume of $(25~h^{-1}{\rm Mpc})^3$. All simulations share the value of these cosmological parameters: $\Omega_{\rm b}=0.049$, $h=0.6711$, $n_s=0.9624$, $w=-1$, $\Omega_K=0$, and $\sum m_\nu=0$ eV.

The simulations have been run with three different codes and, therefore, can be classified into three different suites:

\begin{itemize}

\item \textbf{IllustrisTNG}. The simulations in this suite have been run with the AREPO code \citep{Arepo, Arepo_public} and they employ the IllustrisTNG subgrid physics model \citep{Pillepich_2018, IllustrisTNG_public}.

\item \textbf{SIMBA}. The simulations in this suite have been run with the GIZMO code \citep{Hopkins2015_Gizmo}, and they employ the SIMBA subgrid physics model \citep{SIMBA}.

\item \textbf{Astrid}. The simulations in this suite have been run with the MP-Gadget code \citep{MP-gadget}, and they employ a slightly modified version of the Astrid subgrid physics model \citep{Astrid, Bird_2022}.

\end{itemize}

Each suite contains 1,000 simulations (from the LH set of CAMELS). Each of those simulations has a different value of $\Omega_{\rm m}$, $\sigma_8$, and four astrophysical parameters that control the efficiency of supernova and AGN feedback: $A_{\rm SN1}$, $A_{\rm SN2}$, $A_{\rm AGN1}$, and $A_{\rm AGN2}$. We refer the reader to \cite{CAMELS,CAMELS-Astrid} for further details on the specifics of the astrophysical parameters. We emphasize that the astrophysical parameters have different meanings in each suite due to the different subgrid implementations, and they represent variations relative to the corresponding fiducial models of IllustrisTNG, SIMBA, and Astrid.

The value of these six parameters are arranged in a latin-hypercube with boundaries defined by 

\begin{eqnarray}
    0.1\leq &\Omega_{\rm m}& \leq0.5\\
    0.6\leq &\sigma_8&\leq 1.0\\
    0.25 \leq &A_{\rm SN1}, A_{\rm AGN1}& \leq 4.0\\
    0.5 \leq &A_{\rm SN2}, A_{\rm AGN2}& \leq 2.0~.
\end{eqnarray}
We note that in the case of Astrid, the $A_{\rm AGN2}$ parameter ranges from 0.25 to 4. We also emphasize that all simulations have different values of the initial random seed. In this work, we focus our attention on the $z=0$ snapshots of these simulations.

\subsection{Galaxy properties}
\label{subsec:galaxy_properties}

Halos and subhalos are identified in the simulations using the SUBFIND algorithm \citep{Subfind, Dolag2009}. In this work, we define a galaxy as a subhalo with a stellar mass larger than zero. We follow \citet{Echeverri_2023} and only consider galaxies with stellar masses above $5\times10^8~h^{-1}M_\odot$ to avoid working with small, likely spurious objects. SUBFIND computes many properties for each galaxy, but in this work, we focus our attention on the following 14:

\begin{enumerate}
\item \bm{$M_{\rm g}$}: The subhalo gas mass content, including the circumgalactic medium's contribution.
\item \bm{$M_{\rm BH}$}: The black hole mass of the galaxy.
\item \bm{$M_*$}: The stellar mass of the galaxy.
\item \bm{$M_{\rm t}$}: The total mass of the subhalo hosting the galaxy.
\item \bm{$V_{\rm max}$}: The maximum circular velocity of the subhalo hosting the galaxy:  $V_{\rm max}=\max(\sqrt{GM(<R)/R}$).
\item \bm{$\sigma_v$}: The velocity dispersion of all particles in the galaxy's subhalo.
\item \bm{$Z_{\rm g}$}: The mass-weighted gas metallicity of the galaxy.
\item \bm{$Z_*$}: The mass-weighted stellar metallicity of the galaxy.
\item \bm{${\rm SFR}$}: The galaxy star formation rate.
\item \bm{$J$}: The galaxy's subhalo spin vector modulus.
\item \bm{$V$}: The modulus of the galaxy's subhalo peculiar velocity.
\item \bm{$R_*$}: The radius containing half the galaxy stellar mass.
\item \bm{$R_{\rm t}$}: The radius containing half of the total mass of the galaxy's subhalo.
\item \bm{$R_{\rm max}$}: The radius at which $\sqrt{GM(<R_{\rm max})/R_{\rm max}}=V_{\rm max}$.
\end{enumerate}

For IllustrisTNG simulations, we also consider the following three properties:
\begin{enumerate}
 \setcounter{enumi}{14}
\item \bm{${\rm U}$}: The galaxy absolute magnitude in the U band.
\item \bm{${\rm K}$}: The galaxy absolute magnitude in the K band.
\item \bm{${\rm g}$}: The galaxy absolute magnitude in the g band.
\end{enumerate}

We note that the above three magnitudes are not present in simulations of the SIMBA and Astrid suites because \textsc{SUBFIND} needs some particular properties not stored in those simulations to estimate the magnitudes. We refer the reader to \cite{cosmo1gal} for further details about these properties.

\subsection{Input data}
\label{subsec:data_gen}

The input to our models is a 1-dimensional vector containing the properties of $n$ galaxies, where $n\in[1,10]$. For instance, if we use galaxies from the Astrid suite and set $n=5$, the input vector will contain $5\times14=70$ values. We remind the reader that for IllustrisTNG galaxies, we take 17 properties for each galaxy, while for SIMBA and Astrid only 14 are available. Once the simulation suite and the value of $n$ are chosen, we construct 1,500 1D arrays with the properties of $n$ unique galaxies (i.e., we enforce that the same galaxy cannot appear twice in the same set. It can, however, appear again in a different set). The reason why we take 1,500 arrays is that we have performed several convergence tests, and we find that increasing the number of 1D arrays during training does not yield noticeable improvements in our results.

\subsection{Machine learning techniques}
\label{subsec:ML_techniques}

In this work, we train neural networks to perform likelihood-free inference on the value of 2 cosmological ($\Omega_{\rm m}$ and $\sigma_8$) and 4 astrophysical ($A_{\rm SN1}$, $A_{\rm SN2}$, $A_{\rm AGN1}$, $A_{\rm AGN2}$) parameters.

Our models take as input a 1D vector containing the properties of $n$ galaxies and return $2N_{\rm params}$ numbers, where $N_{\rm params}$ is the number of parameters considered (e.g. $N_{\rm params}=1$ if only inferring one parameter). For each parameter $i$, our models output its marginal posterior mean ($\mu_i$) and standard deviation ($\sigma_i$). This is achieved by minimizing the following loss function:

\begin{eqnarray}
 \mathcal{L} = 
 &&\log \left[ \sum_{ j \in \mathrm{batch} } \left( \theta_{i, j} - \mu_{i, j} \right)^2 \right] + \\ \nonumber
 &&\log \left\{ \sum_{ j \in \mathrm{batch} } \left[ 
 \left( \theta_{i, j} - \mu_{i, j} \right)^2 - \sigma_{i, j}^2 \right]^2 \right\} . \label{eq:loss}
\end{eqnarray}
This loss function guarantees that $\mu,_i$ and $\sigma_i$ represent the parameter's posterior mean and standard deviation $i$ \citep{moment_networks, CMD}. 

Our models use several blocks, each containing a fully connected layer, a LeakyReLU non-linear activation function, and a dropout layer. After the last block, a fully connected layer predicts the network's output. We write our model in PyTorch\footnote{\url{https://pytorch.org/}}. The number of blocks, the number of neurons in the fully connected layers, the learning rate, the weight decay, and the dropout rate are considered hyperparameters.

The value of the hyperparameters is tuned using Optuna\footnote{\url{http://optuna.org/}} \citep{Optuna}, which searches the hyperparameter space for the optimal values of the hyperparameters which minimize the value of the validation loss. We use at least 100 trials and the optimization is done by searching the hyperparameter values that minimize the validation loss. We emphasize that we run Optuna for each different configuration; for instance, when changing the simulation suite or the number of galaxies, we retrain using Optuna to find the best hyperparameters for that case.

To train the models, we first split the simulations into training (850), validation (100), and testing (50) sets. We then construct the input 1D arrays by combining the properties of galaxies from the same simulation. We note that it is important to (1) avoid mixing galaxies from different simulations when combining galaxy properties into the input arrays since different simulations sample different parameter values and (2) avoid having galaxies from the same simulation in different sets (e.g. training and testing), since there could be leakage of information if galaxies from the same simulation are somehow correlated.

\subsection{Performance metrics}
\label{subsec:metrics}

In this work, we use four metrics to quantify the accuracy and precision of our models. To use these metrics, we need to consider that for a given input 1D vector $i$, $\theta_i$ represents the value of the considered parameter, while $\mu_i$ and $\sigma_i$ represent the posterior mean and standard deviation predicted by the network for that parameter. The four statistics we consider are:

\begin{itemize}
    \item \textbf{Root mean squared error:}
        \begin{equation}
            \label{Eqn:Accuracy}
            RMSE=\sqrt{\frac{1}{N}\sum_{i=1}^N (\theta_{i}-\mu_{i})^{2} }~,
        \end{equation}
        where the sum runs over all 1D arrays in the considered test set. Smaller values of the RMSE indicate the model is more accurate.

    \item \textbf{Mean relative error, $\epsilon$:}
        \begin{equation}
            \label{Eqn:Precision}
            \epsilon_i = \frac{1}{N}\sum_{i=1}^N \frac{\sigma_{i}}{\mu_{i}}~,
        \end{equation}
        The mean relative error tells us about the model's precision, with lower values representing, in general, more precise models. The mean relative error does not know anything about the true values. Thus, one can have a very precise but not accurate model. 

    \item \textbf{Coefficient of determination, $R^{2}$} \textbf{:}
        \begin{equation}
            \label{Eqn:R^2}
            R^{2}(\theta_{i},\mu_{i}) = 1 - \frac{ \sum_{i=1}^{n}(\theta_{i}-\mu_{i})^{2}}{\sum_{i=1}^{n}(\theta_{i}-\overline{\theta})^{2}}
        \end{equation}
         where, $\overline{\theta}=\sum_{i=1}^N \theta_{i}$. The $R^2$ quantifies the model's accuracy, with values close to 1 being accurate while values close to zero being poor. 

    \item \textbf{Reduced Chi-squared, $\chi^2$:}
        \begin{equation}
            \label{Eqn:chi^2}
            \chi^{2} = \frac{1}{N}\sum_{i=1}^N \frac{(\theta_{i}-\mu_{i})^{2}}{\sigma_{i}^{2}}~.
        \end{equation}
        We made use of these statistics to quantify the precision of the model error bars (posterior standard deviation). Values close to 1 indicate the size of the errors is appropriate, while values below/above 1 indicate the errors are over/under-predicted.
        
\end{itemize}

\section{Results}
\label{sec:results}

We now present the results of our analysis. We first show the results when training the models using the properties of two galaxies, and then we show the results when considering multiple galaxies.

\subsection{Two galaxies}

\begin{figure*}[t]
\centering
\includegraphics[width=0.90\linewidth]{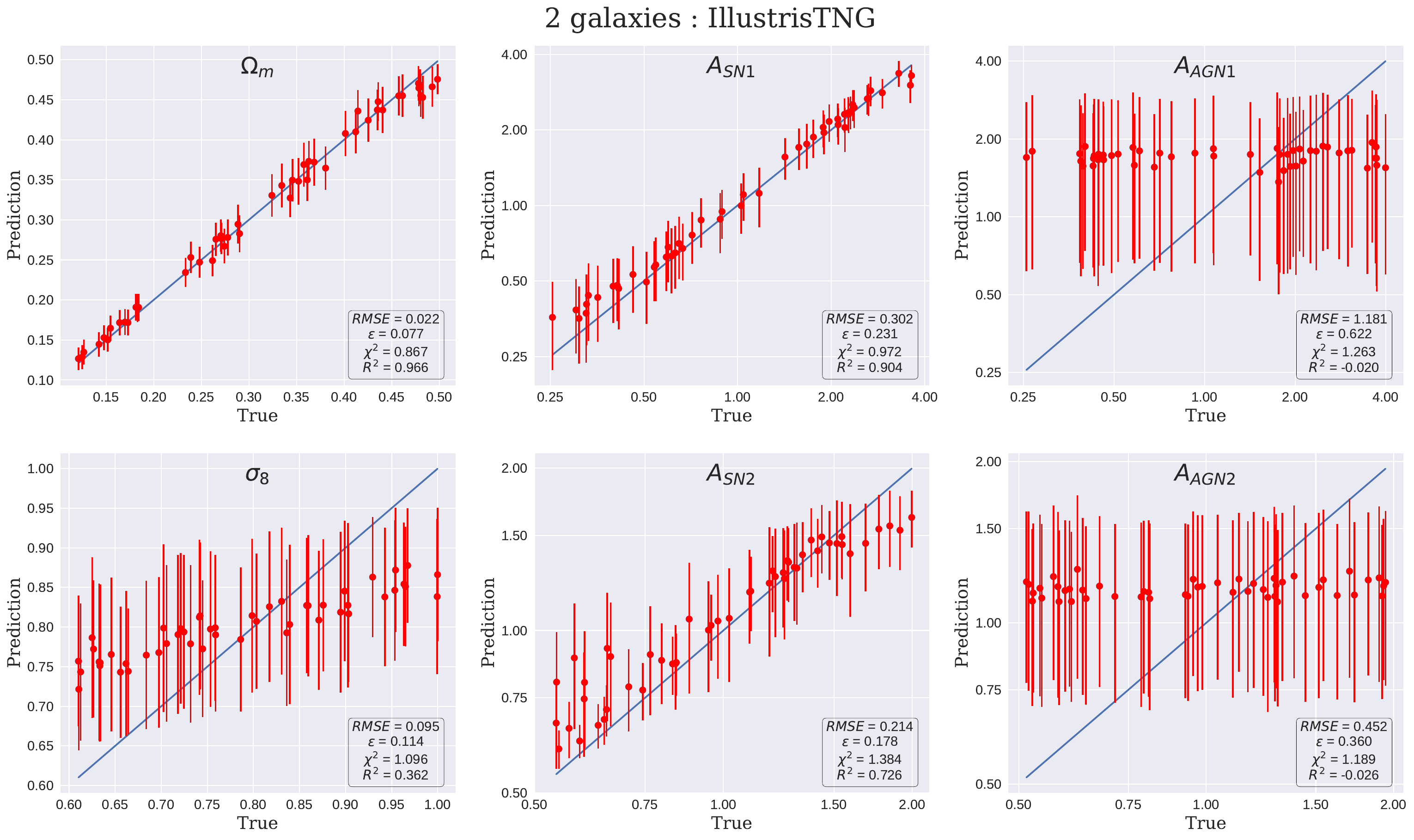}\\
\vspace{0.5cm}
\includegraphics[width=0.90\linewidth]{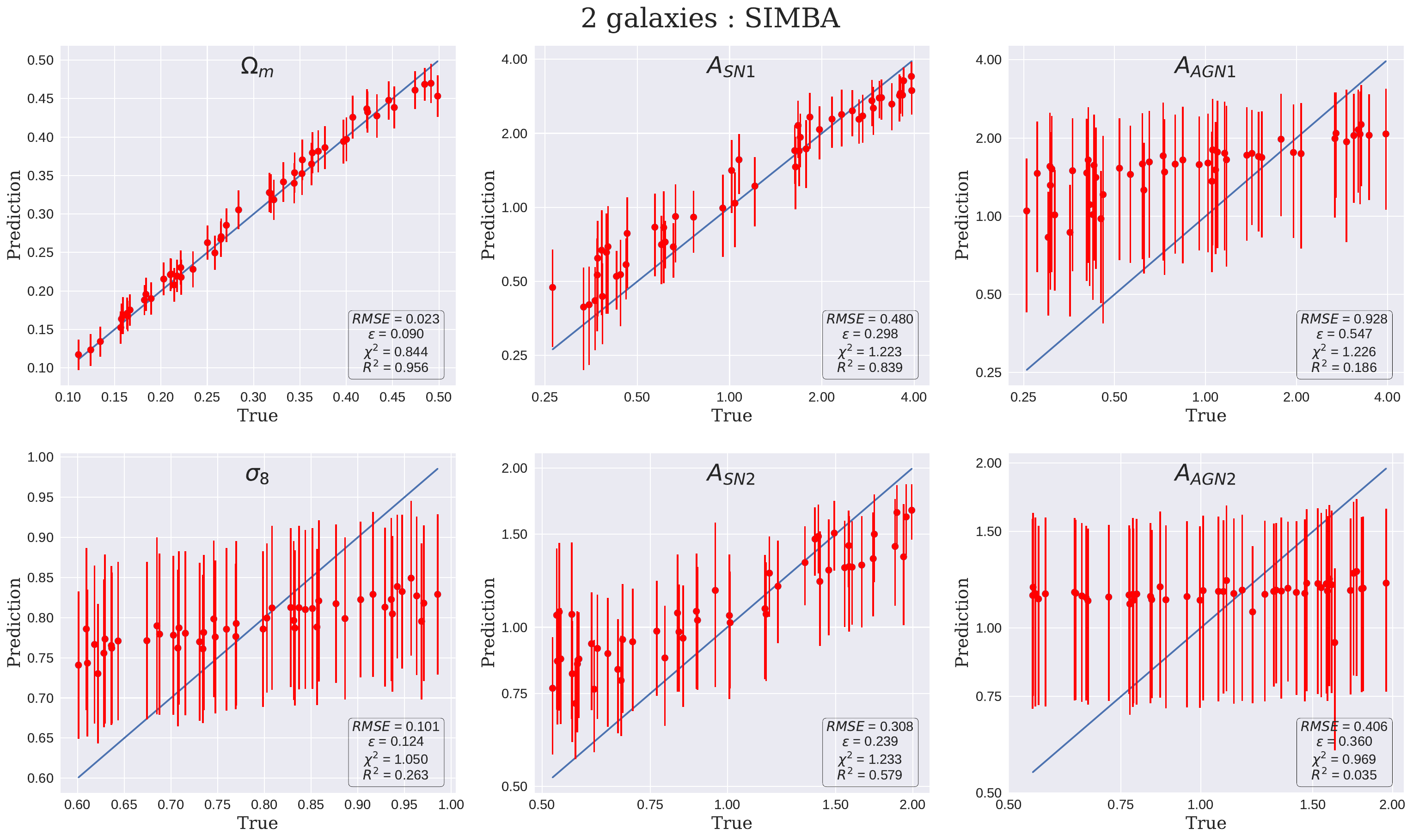}
\caption{We train neural networks to infer the value of the cosmological ($\Omega_{\rm m}$ and $\sigma_8$) and astrophysical ($A_{\rm SN1}$, $A_{\rm SN2}$, $A_{\rm AGN1}$, $A_{\rm AGN2}$) parameters from the internal properties of two random galaxies (without using galaxy positions). Next, from each simulation of the test set, we randomly select two galaxies and test the model on them. We show the results as points with error bars representing the posterior mean and the standard deviation (without making assumptions about the shape of the posterior). As can be seen, the models can precisely infer the value of $\Omega_{\rm m}$ for both IllustrisTNG and SIMBA galaxies and, in some cases, the supernova feedback parameters. The value of $\sigma_8$ and the AGN parameters is poorly predicted in all cases.}
\label{fig:Cosmo2gal_IllustrisTNG_1500}
\end{figure*}

\begin{figure*}[t]
\centering
\includegraphics[width=0.99\linewidth]{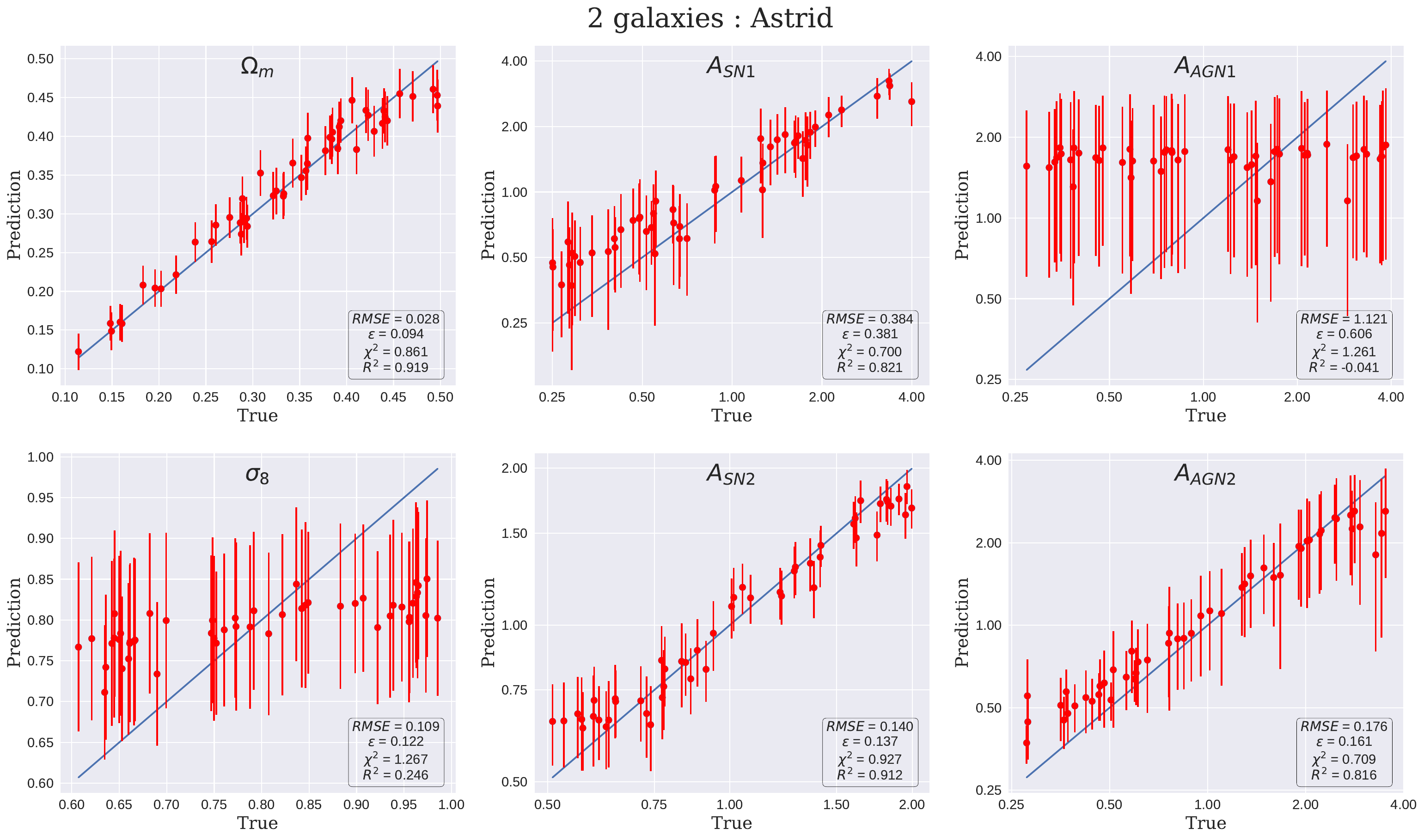}
\caption{Same as Fig. \ref{fig:Cosmo2gal_IllustrisTNG_1500} but for Astrid galaxies.}
\label{fig:Cosmo2gal_Astrid_1500}
\end{figure*}

We train models using 1D arrays that contain the properties of two galaxies. We then test those models on 1D arrays that contain the properties of two galaxies from the test set. We show the results in Figs. \ref{fig:Cosmo2gal_IllustrisTNG_1500} (IllustrisTNG and SIMBA),  and \ref{fig:Cosmo2gal_Astrid_1500} (Astrid). We find that all models can constrain the value of $\Omega_{\rm m}$ accurately with $\{{\rm RMSE}, \epsilon, R^2 \}$ equal to $\{ 0.022, 0.077, 0.966\}$ (IllustrisTNG), $\{ 0.023, 0.090, 0.956\}$ (SIMBA), and $\{ 0.028, 0.094, 0.919\}$ (Astrid). We note that these numbers are better than the ones obtained for a single galaxy; for instance, these metrics are $\{ 0.0365, 0.11, 0.842\}$ when considering one single galaxy from Astrid \citep{Echeverri_2023}. 

On the other hand, $\sigma_8$ remains mostly unconstrained with two galaxies, irrespective of the simulation suite employed, in the same way as our findings for one galaxy \citep{cosmo1gal, Echeverri_2023}. We reach similar conclusions for the AGN parameters of IllustrisTNG and SIMBA simulations. For Astrid, $A_{\rm AGN1}$ remains unconstrained while $A_{\rm AGN2}$ can be inferred with a $\sim 16\%$ precision; a significant improvement from the $\sim24\%$ obtained using one single galaxy \citep{Echeverri_2023}. Finally, all models can infer the supernova feedback parameters with different precisions. We note that in the case of the supernova parameters, we have discarded a very small fraction of galaxies (0.41\% for IllustrisTNG and 0.071\% for SIMBA) since they have unreasonably small widths of the posterior and therefore their $\chi^2$ was really large and affected significantly the reported mean values.

These results show that better constraints on the value of the  parameters can be achieved by using two galaxies instead of one.

\begin{figure*}[t]
\centering
\includegraphics[width=0.99\linewidth]{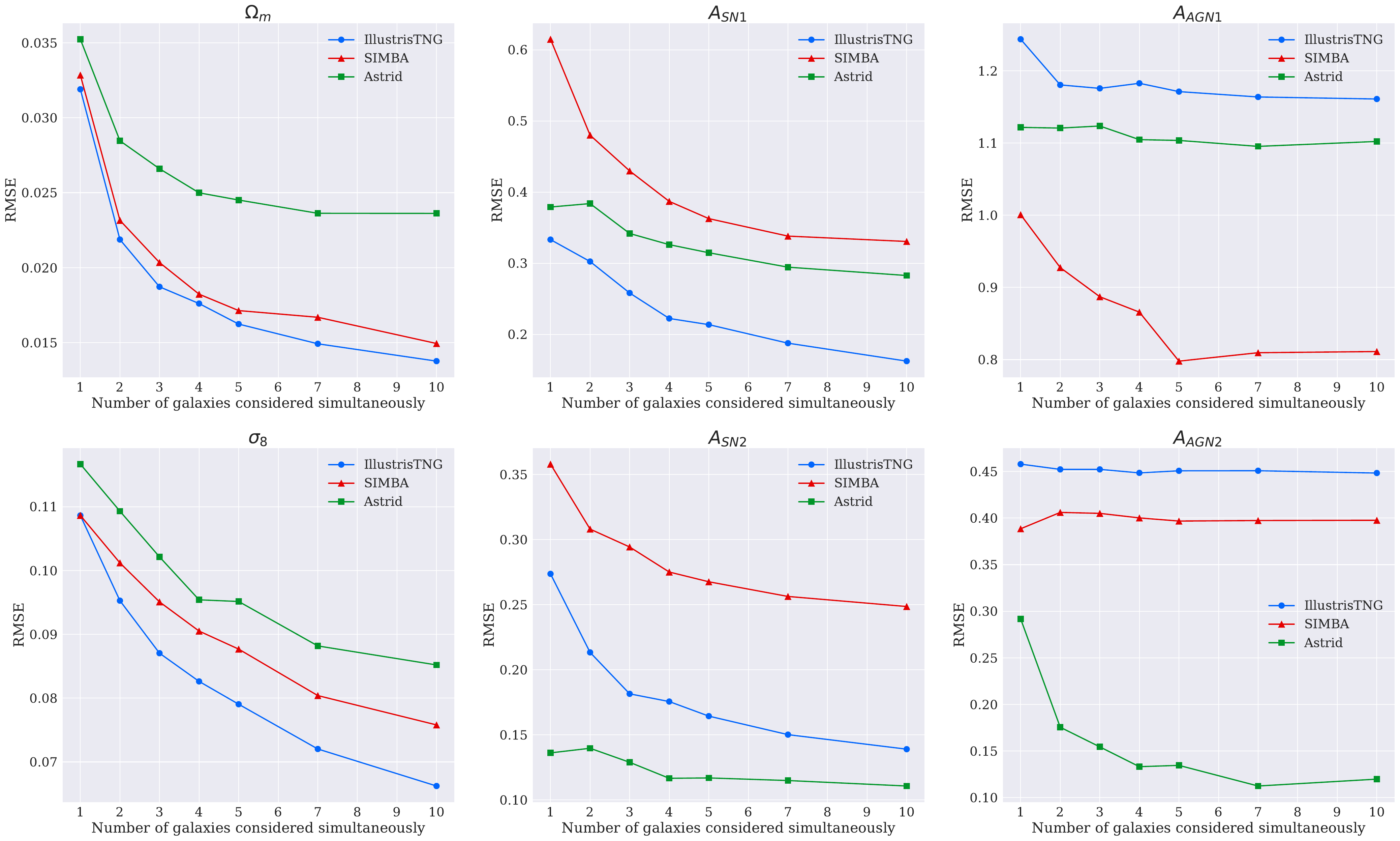}
\vspace{0.5cm}
\includegraphics[width=0.99\linewidth]{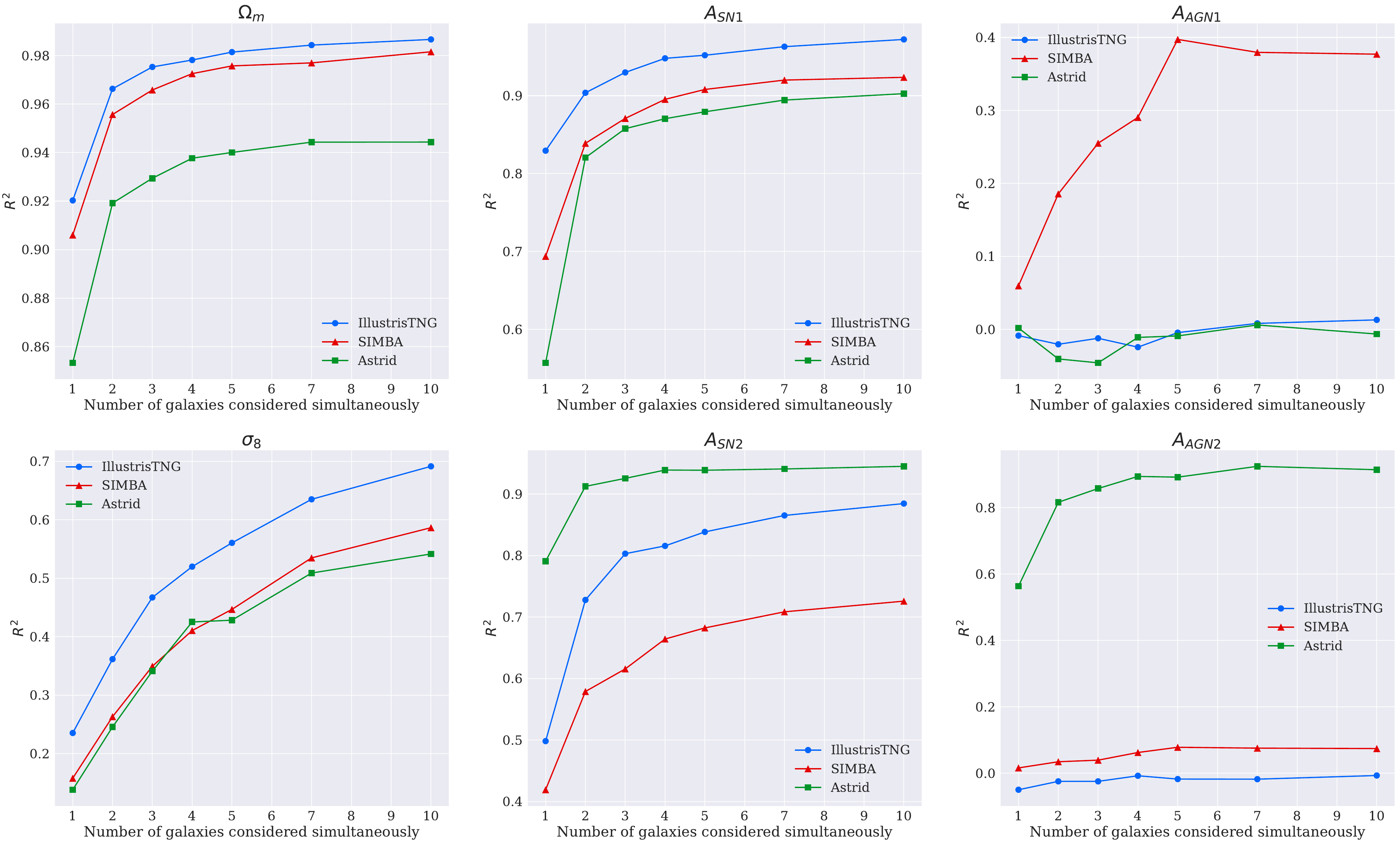}
\caption{We train neural networks to infer the posterior mean and posterior standard deviation of all 6 parameters as a function of the number of galaxies. The top panels show the results for the RMSE, while the bottom panel displays the results for the $R^2$ statistics. In all cases, we show the average results, i.e., for a given simulation, we take 1,500 different combinations and report the mean values. In general, the more galaxies we consider, the tighter the constraints on the parameters. However, there are some cases where constraints saturate, and adding more galaxies does not yield tighter constraints. }
\label{fig:all_accuracy}
\end{figure*}

\begin{figure*}[t]
\centering
\includegraphics[width=0.49\linewidth]{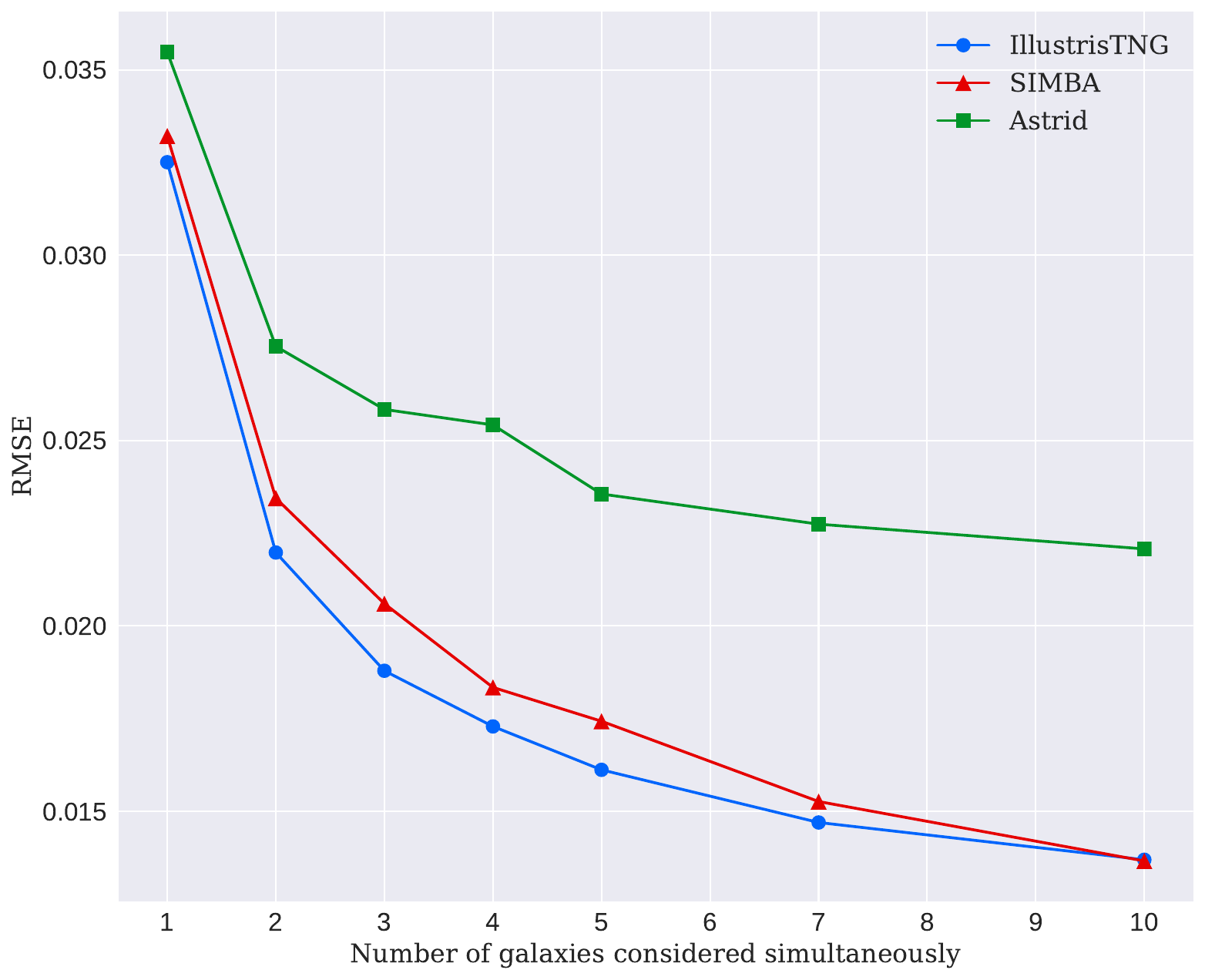}
\includegraphics[width=0.49\linewidth]{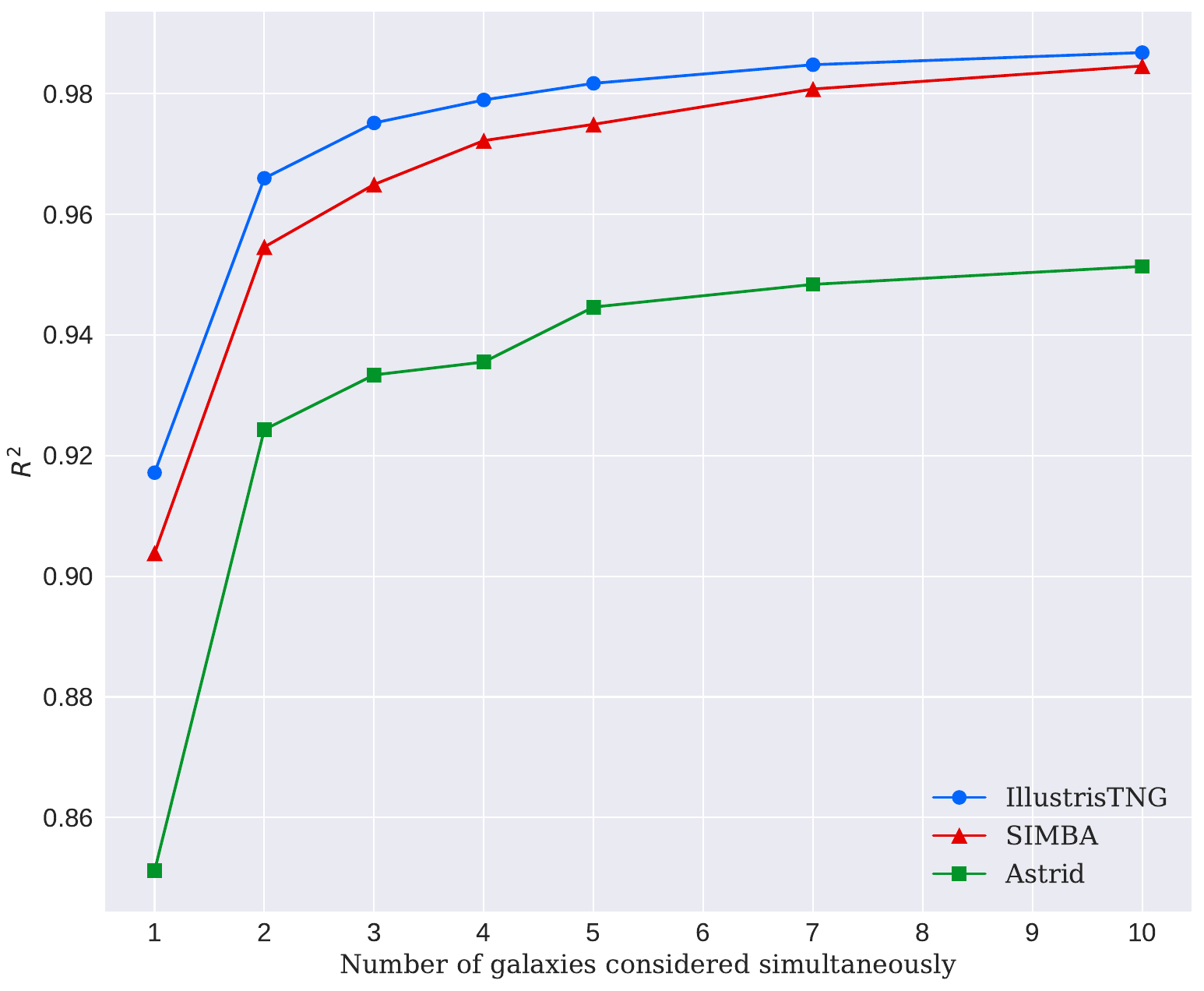}
\caption{We train the models to infer the value of $\Omega_{\rm m}$ alone. The left and right panels show the mean values of the RMSE and $R^2$ as a function of the number of considered galaxies. We find our models perform slightly better when trained to infer $\Omega_{\rm m}$ alone instead of inferring all 6 properties simultaneously. As can be seen, results improve when considering more galaxies, but in the case of Astrid, constraints tend to saturate when using more than $\sim5$ galaxies.}
\label{fig:OnlyOmegam_all_accuracy}
\end{figure*}

\subsection{Multiple galaxies}

Similar to the case of two galaxies, we also carried out the analysis for up to 10 galaxies considered simultaneously. We trained neural networks to perform likelihood-free inferences to estimate the values of the cosmological ($\Omega_{\rm m}$ and $\sigma_8$) and astrophysical ($A_{\rm SN1}$, $A_{\rm SN2}$, $A_{\rm AGN1}$, and $A_{\rm AGN2}$) parameters using data from N galaxies (N goes from 1 to 10) from the IllustrisTNG, SIMBA, and the Astrid suites. Once trained, the model is tested using the galaxies from the test set for each case. Figure \ref{fig:all_accuracy} shows how the prediction RMSE and $R^2$ of the cosmological and astrophysical parameters change as we increase the number of galaxies considered simultaneously. In this case, the quantity reported is the mean value of all galaxies in the test set. In Fig. \ref{fig:Cosmo10gal}, we show the results of training and testing using ten galaxies (this figure is the equivalent to Figs. \ref{fig:Cosmo2gal_IllustrisTNG_1500} and \ref{fig:Cosmo2gal_Astrid_1500}).

We find that, as we consider more galaxies simultaneously, the predicted values of the cosmological parameters $\Omega_{\rm m}$ and $\sigma_{8}$ become increasingly more accurate. In the case of the astrophysical parameters ($A_{\rm SN1}$, $A_{\rm SN2}$, $A_{\rm AGN1}$, and $A_{\rm AGN2}$), their predicted values can either improve or remain the same. The trend in the astrophysical parameters is not the same for all the suites because of the difference in the physical meaning of these parameters in each suite. For instance, the prediction of $A_{\rm AGN2}$ significantly improves when increasing from one galaxy to more than seven for the Astrid model but remains poorly constrained regardless of the number of galaxies in the IllustrisTNG and SIMBA models.

\subsection{Only $\Omega_{\rm m}$}

It is evident from the results discussed until now that the model does an excellent job at predicting the value of $\Omega_{\rm m}$. So, we proceed to train the neural network to predict the posterior mean and standard deviation for only $\Omega_{\rm m}$ instead of all the 6 cosmological and astrophysical parameters. We do this so that the models can focus entirely on minimizing the loss for this parameter, avoiding situations where degeneracies with other parameters can yield suboptimal results for the parameter of interest. 

In this case, these models do a slightly better job at inferring the value of $\Omega_{\rm m}$ compared to results obtained when trained to predict all the 6 cosmological and astrophysical parameters. 

From Fig. \ref{fig:OnlyOmegam_all_accuracy} we see that the neural network becomes increasingly more precise at inferring the value of $\Omega_{\rm m}$ as we increase the number of galaxies considered simultaneously. For SIMBA and IllustrisTNG suites, the RMSE improves by about 55\%, and in Astrid's case, it improves by 37\% as we go from 1 galaxy to 10 galaxies. The right panel of Fig. \ref{fig:OnlyOmegam_all_accuracy} shows the results when considering the $R^2$ statistics instead.

\subsection{Most important features}

\citet{cosmo1gal} carried out a feature importance study that showed that standard feature ranking methods (like computing saliency maps, using SHAP values, or using the inbuilt `feature importance' from scikit-learn) did not yield the important features that the model used to make inferences. This is due to strong internal correlations between galaxy properties, which makes it very difficult for the model to pinpoint the top properties. For that reason, \citet{cosmo1gal} trained a series of gradient-boosted trees models where one feature was discarded at a time. That way, the features could be ranked according to importance, and results were sensitive. \citet{Echeverri_2023} used the same procedure to rank the properties of the Astrid galaxies.  \citet{cosmo1gal} and \citet{Echeverri_2023} found that the 5 most important properties, according to their order of importance, for each of the suites are:
\begin{itemize}
    \item \textbf{IllustrisTNG :} \{$V_{\rm max}$, $M_*$, $Z_*$, $R_*$, $K$\}
    \item \textbf{SIMBA :} \{$V_{\rm max}$, $M_*$, $R_{\rm max}$, $Z_*$, $R_*$\}
    \item \textbf{Astrid :} \{$M_{\rm t}$, $Z_*$, $V_{\rm max}$, $M_{\rm g}$, $M_*$\}
\end{itemize}

With this information on hand, we now ask ourselves whether the constraints we obtain for $\Omega_{\rm m}$ are mostly due to those variables or whether when considering multiple galaxies there may be information coming from other features. 

\begin{figure*}[t]
\centering
\includegraphics[width=0.99\linewidth]{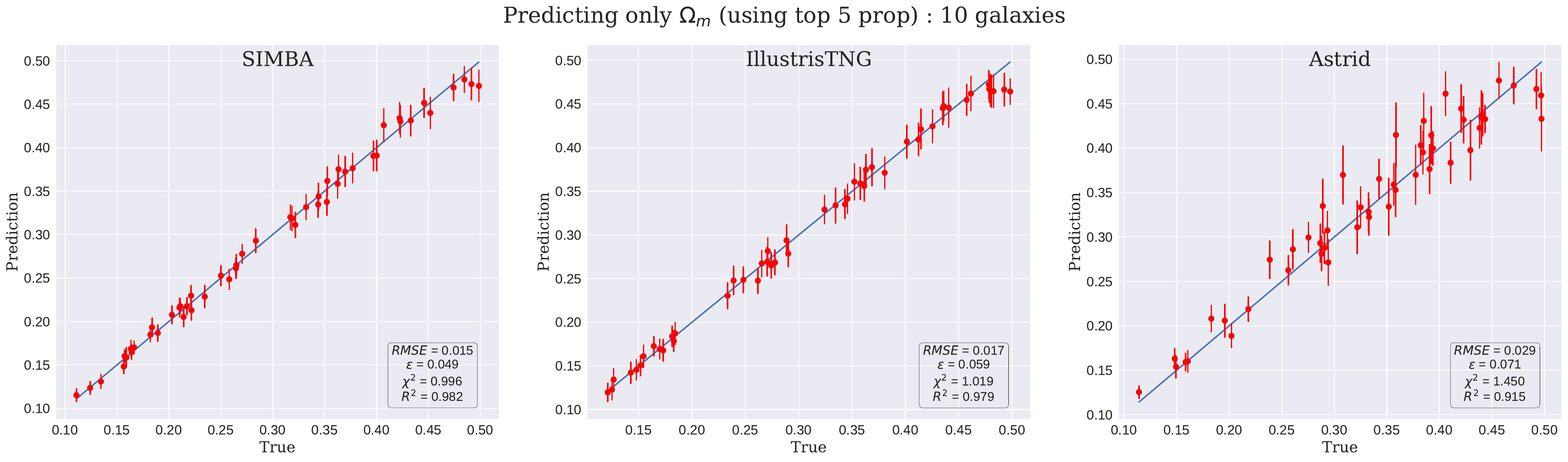}
\caption{We train models to infer the value of $\Omega_{\rm m}$ using the properties of 10 galaxies. However, instead of using all galaxy properties, we made use of the 5 most important properties found when using a single galaxy. The panels show the average results for SIMBA (left), IllustrisTNG (middle), and Astrid (right). We found that constraints using just 5 galaxy properties are similar to the ones obtained using all galaxy properties. This may indicate that the models still extract information from the manifold containing galaxy properties and that information from noisy global quantities (e.g. stellar mass function) is subdominant.}
\label{fig:OnlyOmegam_Cosmo10gal_top5prop}
\end{figure*}

To answer this question, we train models that only use the above galaxy properties but consider multiple galaxies. We show the results in Fig.  \ref{fig:OnlyOmegam_Cosmo10gal_top5prop} for the case of 10 galaxies. As we can see, when predicting only $\Omega_{\rm m}$, the model performs only $\sim6\%$ worse than in the case of SIMBA, and $\sim21\%$ worse than in cases of Astrid and IllustrisTNG, when compared to training with all properties. This is even after removing 12 galaxy properties in the case of IllustrisTNG (9 in the case of SIMBA and Astrid). We thus conclude that most of the information is contained in the most important variables for individual galaxies. We emphasize that this does not mean that the model uses information from individual galaxies and somehow stacks the results. Even using this subset of variables, one can construct noisy estimates of properties, like the stellar mass function, expected to be affected by cosmology \citep{Jo_2022}. Therefore, the source of information may arise from both individual galaxies and collective properties.

\subsection{Robustness}

One of the most important aspects to consider when working with numerical simulations is the robustness of the results. In other words, how well the model behaves when training on galaxies from one galaxy formation model and testing it on galaxies from another galaxy formation models. This aspect has been investigated before in \cite{cosmo1gal}, where it was found that even with a single galaxy, inference of $\Omega_{\rm m}$ from galaxy properties was not robust. This claim was later revisited by \cite{Echeverri_2023}, who found that the lack of robustness was at least partially due to the presence of a small fraction of outliers. 

\begin{figure*}[t]
\centering
\includegraphics[width=0.99\linewidth]{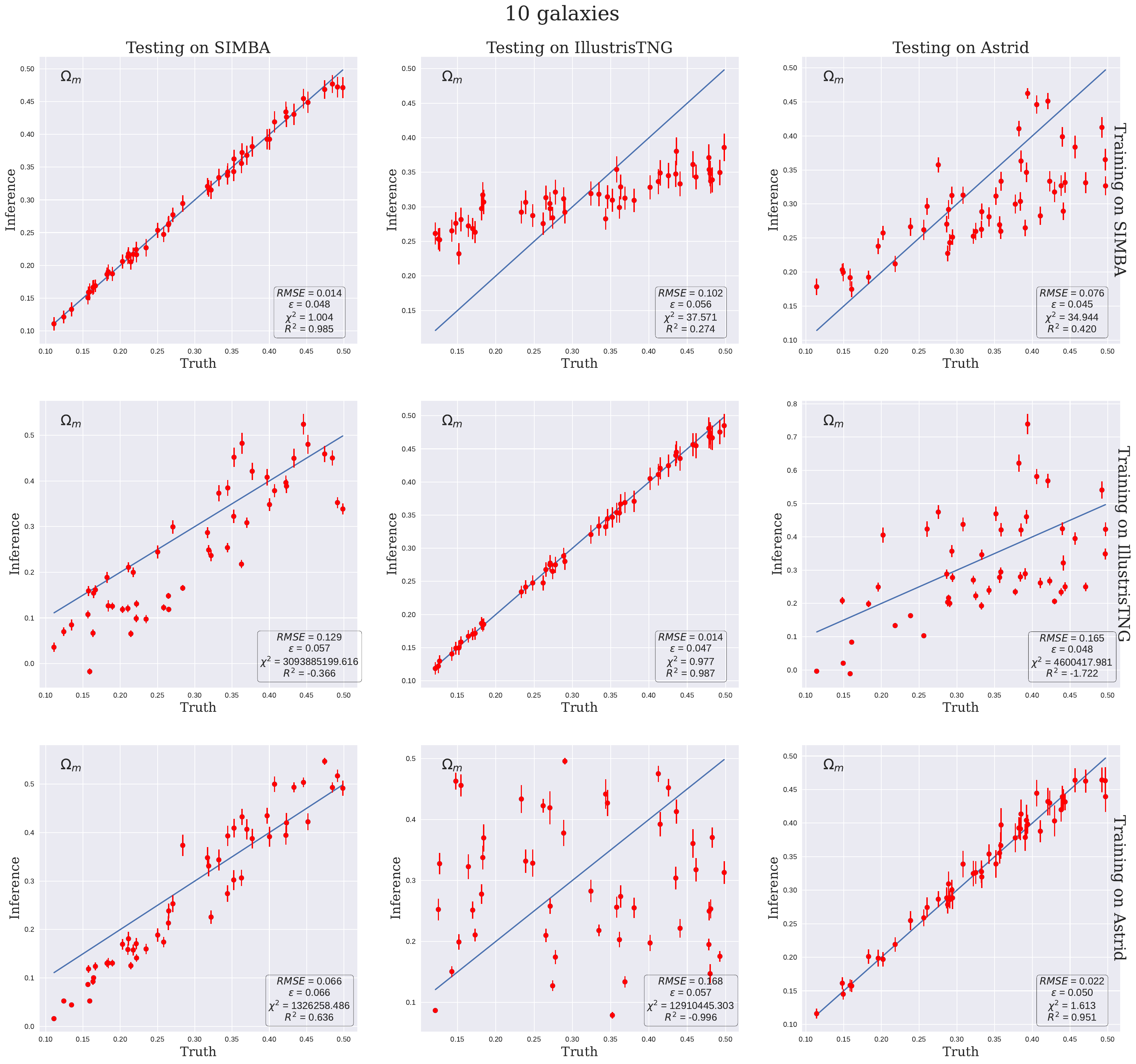}
\caption{Robustness test. We have trained models to infer the value of $\Omega_{\rm m}$ using the properties of 10 galaxies from SIMBA (top row), IllustrisTNG (middle row), and Astrid (bottom row). We then test the models on properties from 10 galaxies from SIMBA (left column), IllustrisTNG (middle column), and Astrid (right column). As can be seen, the models are not robust, and they fail when tested on galaxies from models different from the ones used for training.}
\label{fig:Robustness}
\end{figure*}

In order to verify the robustness of our model, we have considered the case where we train models that use 10 random galaxies produced with a given code and test them using 10 galaxies from another galaxy formation model. We show the results in Fig. \ref{fig:Robustness}. As can be seen, the results are not robust and training models on galaxies from one simulation suite does not yield accurate results when testing on galaxies from another suite. To some extent this is expected since constraints with multiple galaxies are tighter than with a single one and we have not removed, a-priori, outliers from the cross-distributions. We leave for future work to explore strategies designed to increase the robustness of the results following the findings of \cite{Echeverri_2023}.

\section{Summary and Discussion}
\label{sec:summary}

Previous works by \citet{cosmo1gal} and \citet{Echeverri_2023} have pointed out the existence of a tight relation between the properties of individual simulated galaxies and $\Omega_{\rm m}$. The authors interpreted these results as a consequence of the existence of a manifold containing galaxy properties. Under that interpretation, properties of the manifold may change in distinct manners when varying different parameters, allowing the inference of $\Omega_{\rm m}$ from the properties of a single galaxy.

In this work, we have studied whether using the properties of several galaxies can help better constrain the value of the cosmological and astrophysical parameters. To investigate this, we have trained neural networks to perform likelihood-free inferences on the values of the cosmological ($\Omega_{\rm m}$ and $\sigma_8$) and astrophysical ($A_{\rm SN1}$, $A_{\rm SN2}$, $A_{\rm AGN1}$, and $A_{\rm AGN2}$) parameters by using the internal properties of multiple galaxies. We have made use of the properties of galaxies at redshift $z=0$ from the IllustrisTNG, SIMBA, and the Astrid simulation suites from the CAMELS project \citep{CAMELS, CAMELS-Astrid}. We emphasize that our model only uses information from the galaxy properties, not their positions; in other words, the constraints do not incorporate any information from clustering.

We find that the precision of the predictions improves as we increase the number of galaxies. In the case of IllustrisTNG, SIMBA, and Astrid, the RMSE of $\Omega_{\rm m}$ improves by factors of $2.5$, $2.5$, and $1.6$, respectively. In the case of Astrid, we observe a plateau on the constraints when going beyond $\sim5$ galaxies. For IllustrisTNG and SIMBA, the trend indicates that better constraints can be achieved using more than 10 galaxies. When considering the $R^2$ statistics we find that the results tend to saturate when using more than $\sim5$ galaxies. We have trained models to infer the value of $\Omega_{\rm m}$ alone, i.e. without predicting the value of the other parameters. In this case, we find slightly more precise results than when training the models to predict all parameters. The results, shown in Fig. \ref{fig:OnlyOmegam_all_accuracy}, do not change our conclusions.

For $\sigma_8$, we find a steady improvement in the precision of the predictions (both RMSE and $R^2$) as we increase the number of galaxies. We emphasize that our models cannot determine the value of $\sigma_8$ with a single galaxy. The origin of these constraints may arise not from the properties of individual galaxies but from statistics that can be constructed when using multiple galaxies. For instance, the stellar mass function may be sensitive to the value of $\sigma_8$, and a noisy version of it can be constructed when considering multiple galaxies. We thus speculate that the origin of this information may not be related to the manifold hosting the galaxy properties. We note that \cite{Busillo_2023} have obtained cosmological and astrophysical constraints from the properties of a relatively small number of local, star-forming, galaxies.

For the supernova feedback parameters, we also find a consistent improvement in the constraints as we increase the number of galaxies for all suites with the exception of $A_{\rm SN2}$ for Astrid where constraints seem to saturate for more than $\sim5$ galaxies. We believe the explanation may be related to the previous argument, i.e. with multiple galaxies one can construct noisy estimates of global statistical properties that may be sensitive to these parameters, like the stellar mass function or the stellar metallicity relation. However, differently to $\sigma_8$, even with a single galaxy we find some constraining power on the value of these parameters, so it can just be that results are just exploiting that to determine the shape of the manifold better. Both factors likely came into play in this setup.

Finally, we find no constraining power for the AGN parameters for $A_{\rm AGN1}$ for the galaxies in IllustrisTNG and Astrid, and a modest improvement of $\sim20\%$ for SIMBA. On the other hand, for $A_{\rm AGN2}$, the constraints for IllustrisTNG and SIMBA do not improve up to 10 galaxies, while for Astrid there is a significant $\sim2.5\times$ improvement in the RMSE value. We note that AGN feedback is expected to have a larger effect on massive galaxies, so the fact that we choose galaxies randomly (making it more likely to choose small galaxies) can be the reason behind this behavior.

Our results indicate that the models may still be exploiting the information contained on the most important variables used when inferring $\Omega_{\rm m}$ with a single galaxy. In this case, the improvement may be due to a better determination of the galaxy manifold (stacking results for individual galaxies) but also to the impact of cosmology and astrophysics on quantities such as the stellar mass function, where noisy versions of it can be constructed from a set of galaxies. It is however interesting to see that galaxy properties not important for constraints on individual galaxies do not seem to have an impact also when using catalogs.

As expected, our models become more precise but less accurate as we increase the number of galaxies. The reason is that the models are not robust even when considering a single galaxy \citep{cosmo1gal}. However, we note that \cite{Echeverri_2023} found that the models fail, on average, due to the presence of outliers. We thus leave for future work to tackle the robustness of the models for one and multiple galaxies. 

Finally, it is important to compare the results of this work versus those in \cite{Hahn_cosmo1gal}, which are based on the core idea of utilizing the impact of cosmology of galaxy properties. That work provides the first constraints on $\Omega_{\rm }$ and $\sigma_8$ obtained from the photometry alone of thousands of NASA-Sloan Atlas galaxies. In that work, it is found that adding more galaxies improves the constraints, while here, we find that constraints tend to saturate when considering multiple galaxies. However, in \cite{Hahn_cosmo1gal}, the information is not extracted from noiseless galaxy properties but from noisy and dust-attenuated photometry. In that case, even at the level of a single galaxy constraints are poorer than the ones reported here. This is because some information is lost when using photometry instead of galaxy properties. Thus, it is not surprising that stacking thousands of galaxies yields better constraints when using photometry than the ones obtained when using a few galaxies but knowing their properties without errors.

We conclude that better constraints on the value of the cosmological and astrophysical parameters can be obtained by using the properties of multiple galaxies instead of one. In this case, a combination of better knowing the underlying manifold hosting the data and the possibility of constructing noisy estimates of global quantities is behind the performance of our results. It would be interesting to investigate whether some particular combinations of galaxies yield tighter constraints and, therefore, maximize the information content. That selection should also account for the robustness of the model. We leave all this for future work.

\section*{Acknowledgements}

The neural networks have been trained using the GPUs from the Rusty cluster at the Flatiron Institute. We thank Arnab Lahiry, Natali de Santi, and Helen Shao for useful conversations. CC thanks IISER Tirupati for the opportunity to carry out his Master's thesis remotely. The work of FVN is supported by the Simons Foundation. The CAMELS project is supported by the Simons Foundation and the NSF grant AST 2108078. DAA acknowledges support by NSF grants AST-2009687 and AST-2108944, CXO grant TM2-23006X, Simons Foundation Award CCA-1018464, and Cottrell Scholar Award CS-CSA-2023-028 by the Research Corporation for Science Advancement. Details about the CAMELS simulations can be found at \url{https://www.camel-simulations.org}.

\appendix

\section{10 galaxies}

\begin{figure*}[t]
\centering
\includegraphics[width=0.72\linewidth]{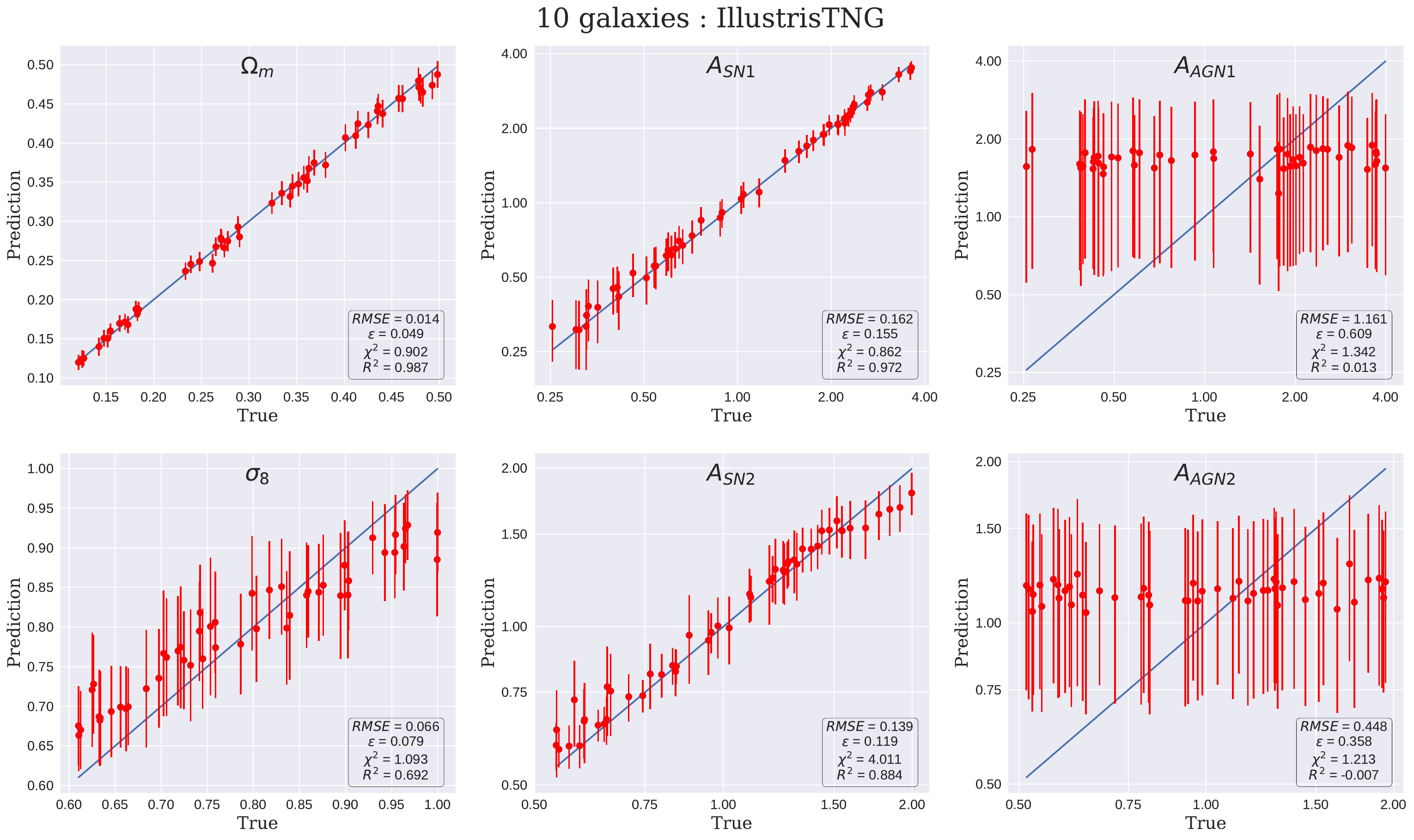}\\
\vspace{0.3cm}
\includegraphics[width=0.72\linewidth]{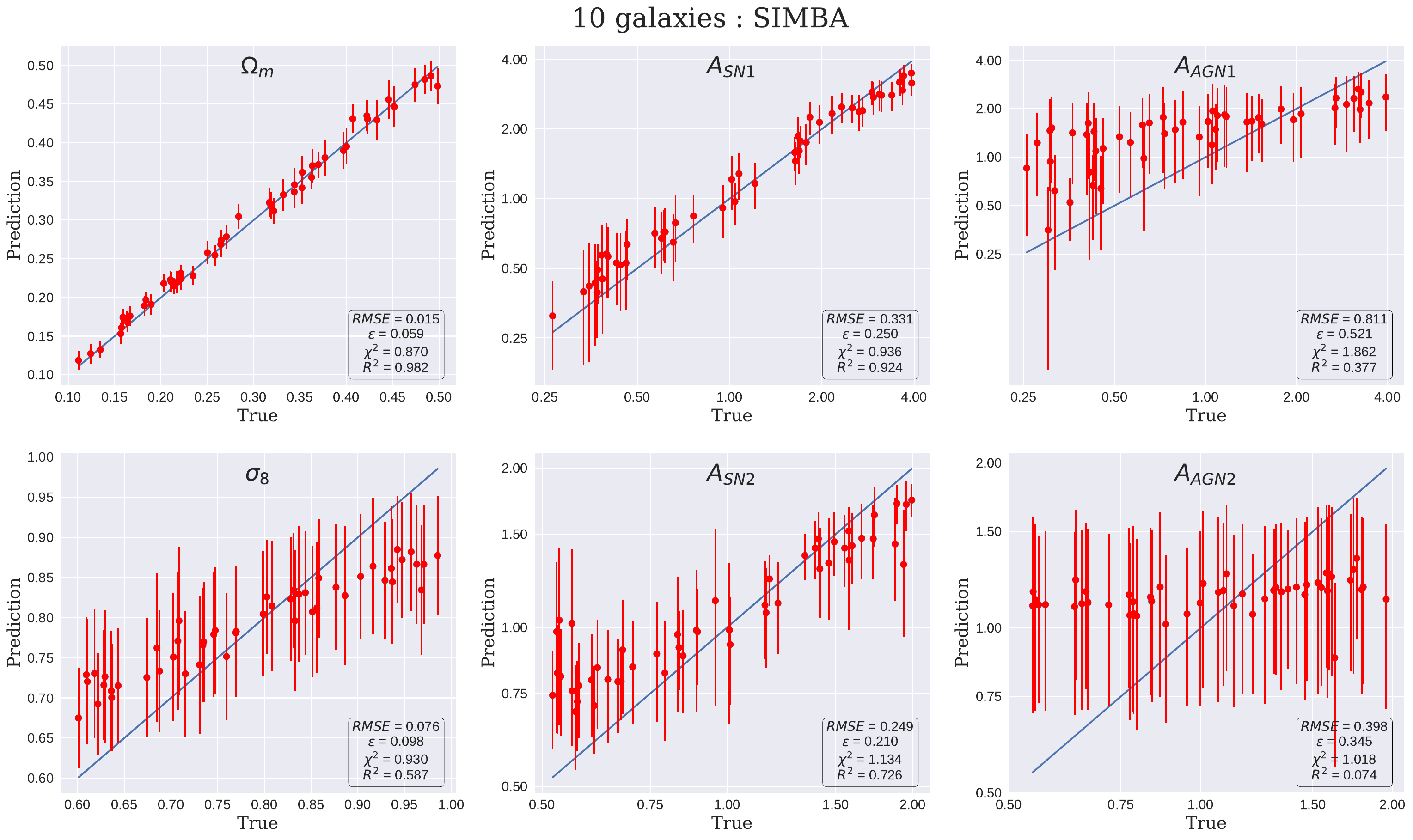}\\
\vspace{0.3cm}
\includegraphics[width=0.72\linewidth]{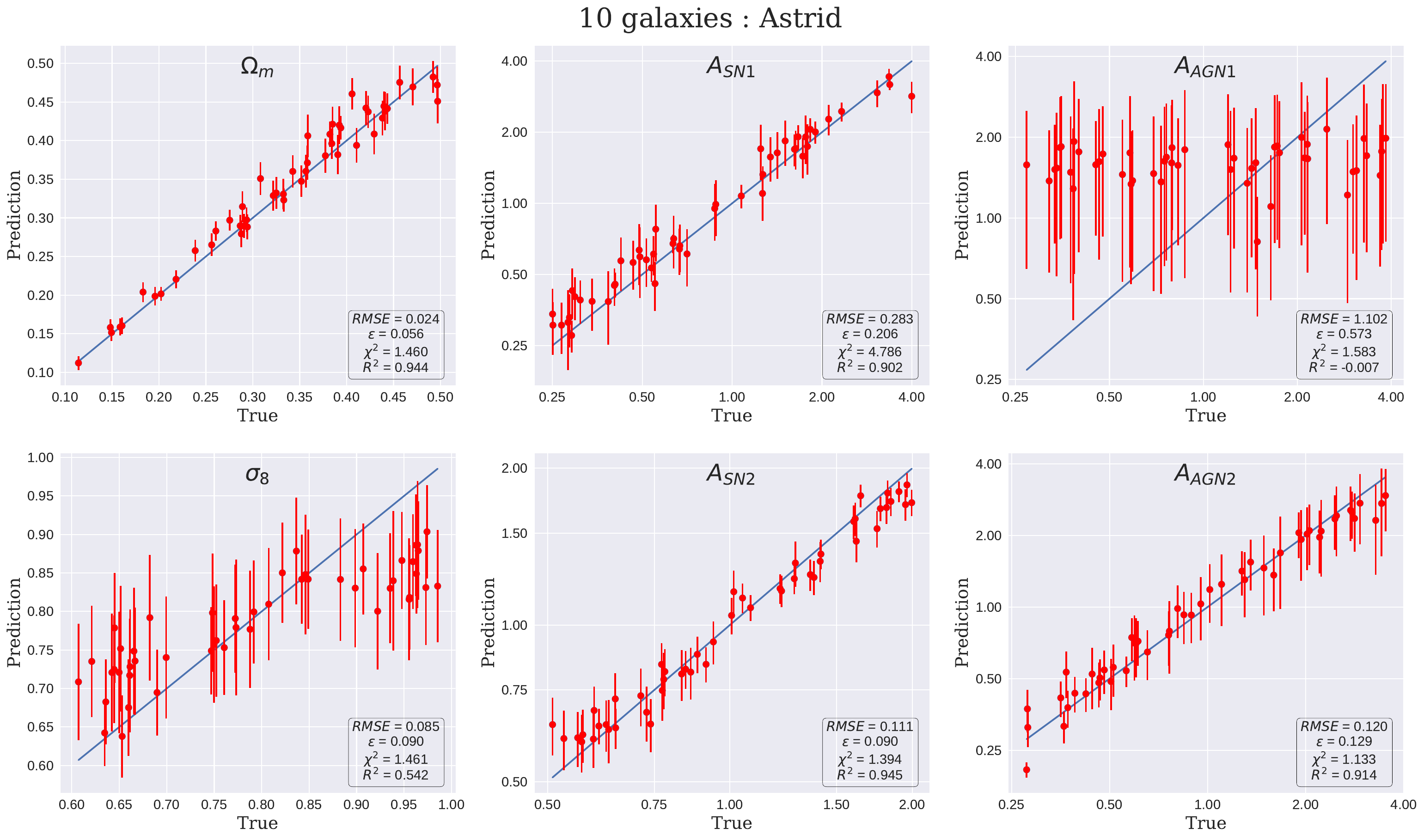}
\caption{Same as Figs. \ref{fig:Cosmo2gal_IllustrisTNG_1500} and \ref{fig:Cosmo2gal_Astrid_1500} but using 10 galaxies instead of 2.}
\label{fig:Cosmo10gal}
\end{figure*}

Fig. \ref{fig:Cosmo10gal} shows the results we obtain when training networks using the properties of 10 galaxies. As can be seen, these results are systematically better than the ones we obtained using two galaxies (see Figs. \ref{fig:Cosmo2gal_IllustrisTNG_1500} and \ref{fig:Cosmo2gal_Astrid_1500}).

\bibliography{references}{}
\bibliographystyle{aasjournal}

\end{document}